\documentstyle[12pt,epsfig]{article}
\newcommand{\be}{\begin{equation}}
\newcommand{\ee}{\end{equation}}
\newcommand{\bear}{\begin{eqnarray}}
\newcommand{\eear}{\end{eqnarray}}
\newcommand{\ba}{\begin{array}}
\newcommand{\ea}{\end{array}}

\newskip\humongous \humongous=0pt plus 1000pt minus 1000pt

\newif\ifdtup

\def\oldreffmt#1{\rlap{[#1]} \hbox to 2\parindent{}}

\def\figfmt#1{\rlap{Figure {#1}} \hbox to 1in{}}

\def\tr{\mathop{\rm tr}}

\def\VEV#1{\left\langle #1\right\rangle}

\def\slash#1{#1\!\!\!/\!\,\,}
\def\beq{\begin{equation}}
\def\eeq{\end{equation}}
\def\bea{\begin{eqnarray}}
\def\eea{\end{eqnarray}}
\def\half{\frac{1}{2}}

\def\bq{\begin{quote}}
\def\eq{\end{quote}}

\def\half{\frac{1}{2}}     
\def \lta {\mathrel{\vcenter
     {\hbox{$<$}\nointerlineskip\hbox{$\sim$}}}}
 
\relax

\newdimen\tdim
\tdim=\unitlength
\def\bar{\overline}

\begin{document}

\pagestyle{empty}
\begin{titlepage}
\def\thepage {}    

\title{  \bf  
The Standard Model in the\\
Latticized  Bulk } 
\author{
\bf Hsin-Chia Cheng$^3$\\[2mm]
\bf  Christopher T. Hill$^1$ \\[2mm]
\bf Stefan Pokorski$^{2}$ \\[2mm]
\bf Jing Wang$^1$ \\ [2mm]
{\small {\it $^1$Fermi National Accelerator Laboratory}}\\
{\small {\it P.O. Box 500, Batavia, Illinois 60510, USA}}
\thanks{e-mail: hcheng@theory.uchicago.edu,
hill@fnal.gov, 
Stefan.Pokorski@fuw.edu.pl, jingw@fnal.gov }\\
{\small {\it $^2$ Institute for Theoretical Physics}}\\
{\small{\it Hoza 69, 00-681, Warsaw, Poland}} \\
 {\small {\it $^3$Enrico Fermi Institute, The University of Chicago}}\\
{\small {\it Chicago, Illinois, 60637 USA}}\\
}

\date{April 12, 2001}

\maketitle

\vspace*{-13.5cm}
\noindent

\begin{flushright}
FERMILAB-Pub-01/053-T \\ [1mm]
April, 2001
\end{flushright}

\vspace*{14.1cm}
\baselineskip=18pt

\begin{abstract}

  {\normalsize
We construct the manifestly gauge invariant
effective Lagrangian in $3+1$ dimensions 
describing the
Standard Model in $4+1$ dimensions, following 
the transverse lattice technique. We incorporate
split generation fermions and we explore
naturalness for two Higgs configurations: a universal Higgs
VEV, common to each transverse brane, and 
 a local Higgs VEV centered on a single brane
with discrete exponential attenuation to other branes,
emulating the split-generation model.  
Extra dimensions, with 
explicit Higgs, do not ameliorate the naturalness
problem. 
} 
\end{abstract}

\vfill
\end{titlepage}

\baselineskip=18pt
\renewcommand{\arraystretch}{1.5}
\pagestyle{plain}
\setcounter{page}{1}


\section{Introduction}

Recently we introduced the low energy effective
Lagrangian of an extra-dimensional Yang-Mills gauge theory 
in which gauge fields, fermions,
and scalars propagate in the bulk \cite{wang1},
\cite{georgi}.  
The idea is to ask  how an experimentalist would describe
the first few KK-modes of, e.g., the gluon, seen in the detector
in an effective Lagrangian in $3+1$ dimensions? 
Hidden local symmetry implies a much larger gauge group than
$SU(3)_{QCD}$
that is spontaneously broken down to $SU(3)_{QCD}$ at low energies,
\cite{hidden}, but how should such a model be wired together
to emulate extra dimensions?
We find that the solution
to this problem is
the transverse Wilson lattice of Bardeen, Pearson
and Rabinovici \cite{trans}.  This leads to  
a local gauge invariant $3+1$ effective Lagrangian 
of the continuum $d+1$ theory with a valid
description of it's attendant
KK-modes in the infra-red, all a consequence of universality.  
The theory is manifestly gauge invariant, renormalizable,
and can be viewed as a new class of models within $3+1$ 
dimensions, with novel
hidden internal symmetries, 
dictated by the imbedding constraints
into extra dimensions.

For example, 
QCD in the $4+1$ bulk
can be described
by a sequence of gauge groups with common coupling,  
$\Pi_{i=0}^{N} SU(3)_i$, with $N$ chiral $(3_i,\bar{3}_{i-1})$ $\Phi_i$
fields connecting the groups sequentially \cite{wang1}.
This can be viewed
as a Wilson action
for a transverse lattice in $x^5$, and is shown 
explicitly to match a compactified  continuum $4+1$ Lagrangian 
truncated in $p^5$ momentum space in axial gauge.   Power-law
running is driven by excitation of
the KK modes with increasing mass scale. 
The
renomalization group is just that of a $3+1$ theory
with many thresholds and can be readily
treated with conventional threshold decoupling techniques \cite{wang1}. 
We find that the transverse lattice is indistinguishable
from the naive $4+1$ running up to very large mass scales. 
We called this an ``aliphatic" model, since
it corresponds to a compactification between
a pair of branes with the field strength $G^a_{\mu5}=0$ on
each brane.\footnote{The name
follows the chemical nomenclature for hydrocarbons; 
aliphatic means ``in a line''}    
The aliphatic model is similar to the
orbifold construction and contains no
undesireable zero-modes, such as massless Nambu-Goldstone
bosons associated with zero-modes of $A_5^a$ components
of the vector potential.

With periodic boundary
conditions the spectrum is changed.
The KK modes are doubled and the overall scale of the
KK masses
increases by a factor of ${2}$.  However, in
the periodic case a zero-mode corresponding
to the $A_5^a$ vector potential component
appears in the spectrum.
As one element
of the present paper, we will explicitly compare
and contrast these two different choices
of boundary conditions, however we will
generally adopt the aliphatic structure throughout
to avoid this spurious Nambu-Goldstone boson.  

Our approach emphasizes that the transverse lattice is
a valid ``completion'' or renormalizable description of
extra dimensions within $3+1$ dimensions.  We argued that
universality allows us to write down any number of
theories that can do this, all yielding the same
infra-red behavior.  The transverse lattice is
optimal, in our opinion, and can be
extended to any number of
dimensions \cite{wang1}.  One can view the transverse lattice
as a Higgs (for links), or gauged chiral Lagrangian,
and any of these descriptions will be equally valid \cite{trans}.
Another example of a high energy completion is the recent
paper \cite{georgi} which proposes a form
of ``Technicolor''
to engineer the effective description of
$4+1$ dimensions.  Note that by opening the closed
moose diagram of ref.\cite{georgi} and removing one
strong condensate, one obtains the aliphatic case,
and removes the unwanted extra Nambu-Goldstone boson.

Our approach can readily be extended to discuss
a wide range of issues.  One can readily construct a
supersymmetric transverse lattice, and one should be
able to describe gravitational KK modes in this approach
as well  \cite{hill}.   Topological and
anomaly questions are also readily 
addressable, and dynamical issues are also better
under control, e.g., for the Nambu-Jona-Lasinio model in
extra dimensions the present approach greatly
simplifies and better defines that analysis \cite{hill}.
This is relevant because extra-dimensions are intrinsically
strongly interacting theories at
some high energy scale, $M_s$,
(this can be seen from perturbative unitarity 
constraints in the $3+1$ theory \cite{wang1}) 
and this may play a fundamental
role in Electroweak Symmetry Breaking (EWSB) \cite{dobrescu}.
The present paper is, in some regards, a warm-up
exercise to return to the study of dynamical (EWSB)
in the transverse lattice formalism.

In the present paper we turn our attention to the
full Standard Model.  Our goal presently is not
ambitious; rather than contructing a
new dynamics for EWSB, we wish to use the usual Higgs mechanism to
describe the EWSB in the full
Standard Model, and to understand the immediate
ramifications of extra-dimensions from the point
of view of the latticized effective Lagrangian.  

This is a transverse  lattice description of a Standard Model
in $4+1$ dimensions in
which the gauge fields and fermions and Higgs all
live in the bulk \cite{dobrescu,dienes,schmaltz}. One
simple and immediate result is that the KK-modes $\gamma_n$, $W_n$ and
$Z_n$ are seen to have a fine-structure in their spectrum
which follows the mass spectrum of the observed Standard Model
$\gamma$, $W$ and $Z$.  

For the description of matter fields,
we exploit the fact that chiral
fermions can always be engineered with arbitrarily massive 
vectorlike KK modes (using the method
as in \cite{kaplan}), so we need keep only the chiral zero-modes.
Indeed, it is an advantage of the $3+1$ formalism that we
can do this; in a sense the chiral generations are put in by hand,
but they can be localized, or split, arbitrarily throughout the bulk.

Presently we will discuss ``split'' inter-generations,
\cite{schmaltz}, \cite{dvali}
but not the more esoteric split intra-generations \cite{schmaltz}.
The latter very interesting
case raises anomaly questions that we have not yet explored, but
which are under current study \cite{hill}.  In the present paper 
we will consider the two cases of (i) a Higgs which develops a common
VEV on all branes, and (ii) the split-generation model in which
we have a localized VEV and generation hierarchy is explained
by the ``distance'' a given generation apperas from the 
localized VEV. This has an elegant
formulation in the present mode, and indeed we
find in the present approach that the split-generation model 
is more general than an extra-dimensional scheme
and may be viewed as a class of generalized
Standard Models in $3+1$ dimensions.

In both Higgs VEV configurations we discuss naturalness.
These two cases are extreme limits on the range of possibilities.
Unfortunately, it appears that extra-dimensions cannot
solve the naturalness problem of the electroweak hierarchy
with a fundamental Higgs.

\section{Standard Model Effective Lagrangian} 

\subsection{Incorporation of QCD; $4+1$ Boundary Conditions vs. 
the $3+1$ Model Structure}

We wish to describe the low energy effective
Lagrangian of the Standard Model
in $4+1$ dimensions using the transverse lattice.  
We begin with the QCD content.
The spectrum of KK modes is sensitive to the structure of
the effective Lagrangian in $3+1$, which
in turn depends upon the global
boundary conditions of the underlying $4+1$ theory.  First
we examine the simplest case, the {\em aliphatic model} corresponding
to a linear system with free boundary conditions \cite{wang1}.
Then we examine the   
{\em periodic model} in which we link the zeroth and $N$th
fields together with one extra link-Higgs field.  
These are distinct global systems with 
characteristically distinct spectra.  Which one occurs
depends upon the 
detailed compactification scheme of nature.

 Consider the pure gauge Lagrangian in $3+1$ dimensions:
\be
{\cal{L}}_{QCD}= -\frac{1}{4}\sum_{i=0}^N G_{i\mu\nu}^a G^{i\mu\nu a} +
\sum_{i=1}^{N} D_{\mu}\Phi_i^\dagger D^{\mu}\Phi_i
\ee
in which we have $N+1$ gauge
groups $SU(3)_i$ and $N$ link-Higgs fields, $\Phi_i$
forming $({3}_{i}, \bar{3}_{i-1})$
representations. The covariant derivative is 
defined as $D_{\mu}= \partial_{\mu} + i
\tilde{g}_3 \sum_{i=0}^{N} 
A_{i\mu}^{a}T_i^{a}$. $\tilde{g}_3$ is 
a dimensionless gauge coupling constant 
that is common to all of
the $SU(3)_i$ local gauge symmetries.
The physical observed low energy QCD coupling will
be $g_3\propto \tilde{g_3}/\sqrt{N+1}$. $T_i^{a}$ are the generators of
the $i$th $SU(3)_i$ gauge symmetry, where $a$ is the color index.
Thus, $[T^i,T^j]=0$ for $i\neq j$; 
$T_i^a$ annihilates a field that is singlet
under the $SU(3)_i$; when the covariant derivative
acts upon $\Phi_i$ we have a commutator of
the gauge part with $\Phi_i$, $T^{a\dagger}_i$ acting on
the left and $T_{i-1}^a$ acting on the right;
the $i$th field strength is $G_{\mu\nu}^{ai} \propto
 \tr T^{ai}[D_\mu,D_\nu]$, etc.

A common renormalizable potential can be constructed for 
each of the link-Higgs fields, 
\be
V(\Phi_j) = \sum_{j=1}^{N} \left[ -M^2 {\rm Tr}(\Phi_j^\dagger\Phi_j) + \lambda_1 {\rm
Tr}(\Phi_j^\dagger\Phi_j)^2 + 
\lambda_2 ({\rm Tr}(\Phi_j^\dagger\Phi_j))^2 + M^{'}( e^{i\theta}\det(\Phi_j)+h.c.) \right],
\label{V} 
\ee
We can always  
arrange the parameters in the potential such that the diagonal
components of each $\Phi_j$ develop a vacuum 
expectation value $v$, and the Higgs and $U(1)$ PNGB are heavy.
Hence, we can arrange that each $\Phi_i$ becomes effectively a
nonlinear-$\sigma$ model field:
\be 
\Phi_i \rightarrow v\exp(i\phi^a_i T^a_i v)
\ee
Thus, the $\Phi_i$ kinetic terms lead to a mass-squared matrix for the gauge
fields:
\be
\sum_{i=1}^{N} \half \tilde{g}_3^2v^2(A^a_{(i-1)\mu} -A^a_{i\mu})^2
\ee
This mass-squared matrix has the structure of a nearest neighbor 
coupled oscillator
Hamiltonian.  It can be written as an $(N+1)\times(N+1)$
matrix sandwiched
between the column vector $A=(A^a_{0\mu}, A^a_{1\mu}, ...,
A^a_{N\mu})$, and it's transpose, as $A^TMA$, where:
\be
\label{mat1}
M =   \half{\tilde{g}_3^2v^2}\left(
\begin{array}{ccccc}
1&-1&0&\cdots&0 \\
-1&2&-1& \cdots&0 \\
0&-1 &2 &\cdots&0 \\
& & & \cdots & \\
0&0&\cdots&-1&1
\end{array} \right).
\ee
We can diagonalize the 
matrix as follows.
The gauge fields $A_{\mu}^{j}$ can be expressed as 
real linear combinations of the
mass eigenstates $\tilde{A}_{\mu}^{n}$ as: 
\be 
A_{\mu}^j = \sum_{n=0}^{N} a_{jn} \tilde{A}_{\mu}^n. 
\label{AA}
\ee
The $a_{jn}$ form a normalized
eigenvector ($\vec{a}_{n}$)
associated with the $n$th $n \neq 0$ eigenvalue 
and has the following components:
\be
\label{2.7}
a_{jn} =\sqrt{\frac{2}{N+1}} \cos{(\frac{2j+1}{2}\gamma_n})\ , \qquad
j = 0, 1,\dots, N, 
\label{anj}
\ee
where $\gamma_n=\pi n/(N+1)$ and $\vec{a}_0 =
\frac{1}{\sqrt{N+1}}(1,1,\cdots 1)$.
The mass terms take the form:
\begin{eqnarray}
{\cal L} _{mass} & = & \half \tilde{g}_3^2v^2\sum_{j=1}^{N} (A_{j-1}-A_j)^2 \\
 & = &  2\tilde{g}_3^2v^2\sum_{n=0}^{N} \sin{(\frac{\gamma_n}{2})}^2
(\tilde{A}^{n})^2. 
\end{eqnarray}
hence the KK tower of masses is:
\be
\label{Mn}
M_n = {2}\tilde{g}_3v \sin \left[ \frac{\gamma_n}{2} \right]
\qquad \gamma_n=
\frac{n\pi}{N+1}\ , \qquad n=0,1,\dots, N. 
\ee
Thus we see that for small $n$ this system has a 
geometrical KK tower
of masses given by:
\be
M_n \approx   \frac{\tilde{g}_3v\pi n}{(N+1)}\qquad \qquad n \ll N
\ee
and $n=0$ corresponds to the zero-mode gluon. 
To match on to the spectrum of the KK modes, we require 
\be
\frac{\tilde{g}_3v\pi}{(N+1)} = \frac{\pi}{R},  
\ee
where $R$ is defined as the size of the $5$th dimension compactified on
the line segment with the boundary condition $G_{\mu 5}=0$ (equivalent to an
orbifold $S_1/Z_2$).   
Hence, the aliphatic system with $SU(3)^{N+1}$ and $N$ $\Phi_i$ provides 
a gauge invariant description 
of the first $n$ KK modes by generating the same 
mass spectrum. 

The zero mode theory is pure QCD with a massless gluon. The
zero-mode trilinear coupling constant 
is $g_3=\tilde{g}_3/\sqrt{N+1}$  \cite{wang1}.
In a geometric picture, the 
aliphatic model corresponds to a
``transverse lattice'' description of a full $4+1$
gauge theory \cite{trans}, where the $4+1$ theory
is compactified between two parallel branes at $x^5=0$
and $x^5=R$ and the boundary conditions
on the branes are $G_{\mu 5}^a = -G_{5 \mu }^a =0$.
These boundary conditions insure that no vector gauge
invariant field strength is ``observable'' on the branes.
There is no $A_5^a$ zero-mode (all of the $N$ link-Higgs
chiral fields have been eaten to provide longitudinal components
to the massive KK mode gluons).

Of course, we can always make a periodic extension
of the interval $[0,R]$. 
This 
leads to a Lagrangian in which we have $N+1$ branes,
hence $N+1$ $SU(3)_i$ as before, but
now, $N+1$ linking $\Phi_i$ Higgs fields, 
\be
{\cal{L}}= -\frac{1}{4}\sum_{i=0}^N G_{i\mu\nu}^a G^{i\mu\nu a} +
\sum_{i=0}^{N} D_{\mu}\Phi_i^\dagger D^{\mu}\Phi_i
\ee
We now have the additional
$\Phi_0$ which is a $(\bar{3}_0,3_N)$ representaion linking
the first $SU(3)_0$ gauge group to the last $SU(3)_N$.
The resulting gauge field mass-squared term becomes:
\be
\sum_{i=1}^{N+1} \half \tilde{g}_3^2v^2(A^a_{(i-1)\mu} -A^a_{i\mu})^2
\ee
where we identify $A^a_{(N+1)\mu}\equiv A^a_{(0)\mu}$.
Thus, the mass-squared matrix is now:
\be
M =   \half{\tilde{g}_3^2v^2}\left(
\begin{array}{ccccc}
2&-1&0&\cdots&-1 \\
-1&2&-1& \cdots&0 \\
0&-1 &2 &\cdots&0 \\
& & & \cdots & \\
-1&0&\cdots&-1&2
\end{array} \right).
\ee
The diagonalization is now done with
a complex representation (suppressing
gauge and Lorentz indices; consider $N$
even):
\be 
A_{\mu}^j = \sum_{n=-N/2}^{N/2} a_{jn} \tilde{A}_{\mu}^n. 
\label{AA}
\ee
where now:
\be
\label{2.17}
a_{jn} = \frac{1}{\sqrt{N+1}}\exp\left(i 2\pi \frac{ n j}{N+1}\right) , \qquad
j = 0, 1,\dots, N, 
\ee
Note with this definition $A_j$ is periodic, $A_{(N+1)}=A_0$.
Reality of $A_i$ dictates that $\tilde{A^n} = \tilde{A}^{-n}{}^*$.
One thus obtains for the mass matrix:
\be
2{\tilde{g}_3^2v^2}\sum_{n=-N/2}^{N/2}  \sin^2\left(\frac{\pi n}{N+1} \right) 
|\tilde{A^n}|^2
\ee
The spectrum is now:
\be
2{\tilde{g}_3v}\sin\left(\frac{\pi n}{N+1} \right) \qquad n=0,1,2, ... , N/2
\ee
We now require: 
\be
\frac{\tilde{g}_3v}{(N+1)} = \frac{1}{R}. 
\ee
Hence, the periodic system with $SU(3)^{N+1}$ and $N+1$ $\Phi_i$ provides 
a gauge invariant description of the first $n$ doubled
KK modes, generating the same 
mass spectrum as in the aliphatic case up to an overall scale
factor of ${2}$.  (Note that 
if $N$ were odd the spectrum would include
an additional singlet level with $n=(N+1)/2$).
There remains the zero-mode in the spectrum
$n=0$, which is a singlet since
the reality condtion  $\tilde{A^n} = \tilde{A}^{-n}{}^*$ 
inplies that  $\tilde{A^0}$ is real.  However, every
nonzero $n$ corresponds to a degenerate doublet of
levels.  

The zero-mode theory of periodic
boundary conditions contains QCD with a massless
gluon and a coupling constant $g_3 = \tilde{g}_3/\sqrt{N+1}$.
Now, however, there is an additional component:  Since
we added one extra link-Higgs there is a zero-mode chiral field
$\phi_0^a$ which is not part of the normal low
energy spectrum of QCD.
This field is a color-octet massless Nambu-Goldstone boson
(NGB) mode. It would bind
with $q\bar{q}$ and with itself
to produce exotic mesons. Most exotic would be 
a boundstate of a gluon and $\phi^a$.  These exotic states might
be heavy, and could decay quickly to normal hadrons,
so it is unclear whether they are
ruled out.  In the case of the electroweak part
of the Standard Model similar objects would also
occur as light Nambu-Goldstone bosons, and are likely
problematic. 

Since our present goal is to construct a low energy
model that is the minimal Standard Model, we are therefore compelled
to use the aliphatic boundary conditions  
to remove these  NGB's.
Henceforth, throughout the remainder of the paper we
will use the aliphatic constructions with $N+1$ gauge
fields and $N$ link-Higgs fields. 

\subsection{Incorporating $SU(2)_L\times U(1)_Y$}

We now
consider the pure gauge Lagrangian in $3+1$ dimensions:
\be
{\cal{L}}_{ew}= -\frac{1}{4}\sum_{i=0}^N F_{i\mu\nu}^a F^{i\mu\nu a} 
-\frac{1}{4}\sum_{i=0}^N F_{i\mu\nu} F^{i\mu\nu}
+\sum_{i=1}^{N} D_{\mu}\Phi'{}_i^\dagger D^{\mu}\Phi'_i
 +\sum_{i=1}^{N} D_{\mu}\phi_i^\dagger D^{\mu}\phi_i
\ee
Here we have $N+1$ copies
of the $SU(2)_L\times U(1)_Y$ 
electroweak Standard Model.  Thus the
gauge group is $\Pi_{i=0}^N SU(2)_{iL}\times U(1)_{iY}$
where
$F_{i\mu\nu}^a$  $(F_{i\mu\nu})$ is the
$SU(2)_{iL}$ ($U(1)_{iY}$) field strength. 
The $N$ $\Phi'_i$ and $\phi_i$ are elementary scalars.  
The $\Phi'_i$ carry $SU(2)$ charges $(\half_i,\half^C_{i-1})$,
where ${}^C$ denotes charge conjugation, and the 
$\phi_i$ carry weak hypercharges $(Y_i, -Y_{i-1})$.  These 
fields correspond to the links
of a transverse Wilson lattice in the fifth dimension,
$x^5$.   

Note that we will ultimately specify
the $\phi$ charges
to be given by $Y_i=Y=1/3$ throughout. We must
choose $\phi_i$ to carry less than the
smallest common unit of the weak hypercharge of all
components of the theory. This serves the purpose of 
constructing the fermion links, as in mass-mixing
operators required for the CKM matrix, out
of polynomial operators involving $\phi^p$,
not allowing fractional powers, $p$. We cannot strictly 
use a product link, 
$\tilde{\Phi} =\Phi'\phi$, which is a slight departure from the pure
transverse lattice.  In what immediately follows we will
write $Y$ as a generic parameter.

We arrange potentials for the $\Phi'_i$ and $\phi_i$ so
they each acquire VEV's independent of $i$.
Hence, we can again arrange that each field becomes effectively a
nonlinear-$\sigma$ model:
\be 
\Phi'_i \rightarrow v_2\exp(i\phi^a_i\tau^a/2v_2)\qquad
\phi_i \rightarrow \frac{v_1}{\sqrt{2}}\exp(i\phi_i/v_1)
\ee
Thus, the $\Phi'_i$ and $\phi_i$ kinetic terms lead to a mass-squared matrix 
for the $SU(2)$ and $U(1)$ gauge
fields:
\be
\sum_{i=1}^{N} \half \tilde{g}_{2}^2v_2^2(A^a_{(i-1)\mu} -A^a_{i\mu})^2
+ \sum_{i=1}^{N} \half \tilde{g}_1^2v_1^2Y^2(A_{(i-1)\mu} -A_{i\mu})^2
\ee
 The gauge fields $A_{\mu}^{j}$ can 
 again be expressed as linear combinations of the
mass eigenstates $\tilde{A}_{\mu}^{n}$ as: 
\be 
A_{\mu}^j = \sum_{n=0}^{N} a_{jn} \tilde{A}_{\mu}^n. 
\label{aa}
\ee
with (in the aliphatic case):
\be
a_{jn} =\sqrt{\frac{2}{N+1}} \cos{(\frac{2j+1}{2}\gamma_n})\ , \qquad
j = 0, 1,\dots, N, 
\ee
where $\gamma_n=\pi n/(N+1)$.
The mass eigenvalues are:
\be
\label{Mn}
M^{(2)}_n = {2}\tilde{g}_2v_2 \sin \left[ \frac{\gamma_n}{2} \right]
\qquad 
M^{(1)}_n = {2}\tilde{g}_1v_1 Y \sin \left[ \frac{\gamma_n}{2} \right]
\qquad \gamma_n=
\frac{n\pi}{N+1}\ , \qquad n=0,1,\dots, N. 
\ee
Thus we see that for small $n$ this system has a KK tower
of masses given by:
\be
M^{(2)}_n \approx   \frac{\tilde{g}_2v_2\pi n}{(N+1)};\qquad
M^{(1)}_n \approx   \frac{\tilde{g}_1v_1Y \pi n}{(N+1)}
\qquad \qquad n \ll N
\ee
and $n=0$ again corresponds to the zero-mode gauge fields. 

To match on to the spectrum of the KK modes, we require 
\be
\frac{\tilde{g}_2v_2}{(N+1)} = \frac{\tilde{g}_1v_1Y }{(N+1)} = \frac{1}{R}. 
\ee
The KK modes should have common
values owing to geometry. Thus we require for matching:
\be
\frac{v_2}{v_1} = \frac{\tilde{g}_1Y}{\tilde{g}_2} = Y\tan\theta_W
\ee
This corresponds to
an aliphatic system with $SU(2)_L^{N+1}\times U(1)^{N+1}$ and $N$ $\Phi'_i$
and $\phi_i$ providing 
a gauge invariant description of the first $n$ KK modes.

The zero modes of this pure gauge theory
are described by the effective Lagrangian in $3+1$ dimensions:
\be
{\cal{L}}_{gauge}= -\frac{1}{4} F_{\mu\nu}^a F^{\mu\nu a} 
-\frac{1}{4}F_{\mu\nu} F^{\mu\nu}
\ee
where 
$F_{\mu\nu}^a$  $(F_{\mu\nu})$ is the
$SU(2)_{L}$ ($U(1)_{Y}$) field strength. 
The physical
$SU(2)_L$ ($U(1)_{Y}$) gauge coupling constant 
is $g_2 \equiv  \tilde{g}_2/\sqrt{N+1}$,
($g_1 \equiv \tilde{g}_1/\sqrt{N+1}$) 
a consequence of using the expansion of 
eq.(\ref{aa}). The fact that the
physical coupling constants are suppressed by $\sim 1/\sqrt{N}$
is just the classical volume supression 
of the coupling in the $4+1$ dimensional
theory.

\section{Incorporating Electroweak Higgs Fields}

We now introduce $N+1$ Higgs fields, $H_i$
each transforming as $\left(\half_i\right)$ under
$SU(2)_i$ (and singlet under $SU(2)_j$
$j\neq i$), and with weak hypercharges $Y_i=1$ 
(and $Y_j=0$ $j\neq i$).
The Lagrangian for the Higgs fields is:
\be
\label{higgs1}
{\cal{L}}_{Higgs} =
 \sum_{i=0}^{N} (D_\mu H_i)^\dagger (D^\mu H_i)
 - M_0^2|H_{i+1} -(\Phi'_{i+1}\phi_{i+1}^3/v_1^3v_2)H_i|^2 - V(H_i)
\ee
where we identify  $H_{N+1}=0$ in
the aliphatic case.  Here we have chosen $Y=1/3$,
and thus the $\phi^3$ link appears.
Note that the second term is a latticized covariant derivative
in the $x^5$ direction.  Purely from the
point of view
of the $3+1$ theory it is advantageous to rewrite
eq.(\ref{higgs1}) as:
\bea
\label{higgs2}
{\cal{L}}_{Higgs} & = &
 \sum_{i=0}^{N} \left[(D_\mu H_i)^\dagger (D^\mu H_i)
 - 2M_0^2|H_{i}|^2 +   \lambda'(H_{i+1}(\Phi'_{i+1}\phi_{i+1}^3)H_i^{\dagger}
 + h.c.) - V(H_i) \right] \nonumber \\
& & + M_0^2|H_{0}|^2 + M_0^2|H_{N}|^2
\eea
The last terms take care of the difference
between $H_0$, $H_N$ and $H_i$ in the aliphatic case.  
Note that $\lambda' = M_0^2/v_1^3v_2$.
The theory now appears
as a conventional $3+1$ multi-Higgs
model with a system of mass-terms and higher
dimension interactions 
with the link-Higgses.

First we
ignore the Higgs potentials, and we gauge
away the chiral field components, so $\Phi'_i= v_2$ and
$\phi_i= v_1$. 
We thus have in eq.(\ref{higgs1}) the nearest neighbor
mass terms:
\be
{\cal{L}}_{Higgs} =
 -\sum_{n=1}^{N} 
 M_0^2|H_{i-1} -H_i|^2 
\ee
which leads to the spectroscopy:
\be
\label{mhn}
M^2_n = 4M_0^2 \sin^2 \left[ \frac{\gamma_n}{2} \right]
 \qquad n=0,1,\dots, N. 
\ee
Matching onto the spectrum of the KK modes requires: 
\be
 \frac{M_0}{(N+1)} = \frac{1}{R}. 
\ee
The eigenfields are given by:
\be 
H^j = \sum_{n=0}^{N} a_{jn} \tilde{H}^n. 
\label{hh}
\ee
with the $a_{jn}$ as in eq.(\ref{2.7}).

We now incorporate the  Higgs potentials.
We consider presently a universal Higgs potential common
to each brane $i$ (we will consider a nonuniversal
configuration in the subsequent section):
\be
V(H_i) = -\tilde{m}^2 H_i^\dagger H_i + 
\frac{\tilde{\lambda}}{2} (H_i^\dagger H_i)^2
\ee
The presence of the Higgs potential adds a common
mass term $-\tilde{m}^2 \sum H_i^\dagger H_i$ 
to each of the $H_i$ in the Lagrangian. This modifies
the eigenvalues:
\be
\label{mhn}
M^2_n = {4}M_0^2 \sin^2 \left[ \frac{\gamma_n}{2} \right] - \tilde{m}^2
 \qquad n=0,1,\dots, N. 
\ee
We see that $-\tilde{m}^2$ is the mass for the zero mode.
Hence the zero-mode Lagrangian corresponds to the Standard
Model with a tachyonic Higgs of negative mass-squared  $-\tilde{m}^2$.   

Let us go to mass eigenbasis and
truncate on the zero-mode. Hence
 the zero-mode Higgs potential is:
\be
V(\tilde{H}_0) = -\tilde{m}^2  \tilde{H}_0^\dagger \tilde{H}_0
+ \frac{\tilde{\lambda}}{2(N+1)} (\tilde{H}_0^\dagger \tilde{H}_0)^2
\ee
Notice the large suppression factor of the quartic interaction
term, a consequence of the normalization of
the zero-mode component of the Higgs field. 
This may be interpreted as the volume supression of
the quartic coupling constant in the extra-dimensional
theory.  Thus, we define the
low energy physical quartic coupling as $\lambda =
\tilde{\lambda}/{(N+1)}$.
The VEV of the zero mode Higgs, $\VEV{\tilde{H}_0} = (v_0, 0)^{T}$, 
thus becomes  $v_0^2 = \tilde{m}^2/\lambda = (N+1)
\tilde{m}^2/\tilde{\lambda}$.  
Substituting the zero-mode Higgs field with VEV,
the zero-mode Higgs boson kinetic term becomes:
\be
{\cal{L}}_{Higgs} =
 \sum_{j=0}^{N} (D_\mu H_j)^\dagger (D^\mu H_j)
 \rightarrow 
\frac{1}{(N+1)} \sum_{j=0}^{N} \left|(i\tilde{g}_2 A_{j, \mu} ^a\frac{\tau^a}{2}+i\tilde{g}_1 A_{j,\mu} \frac{Y}{2})
 \left( \begin{array}{c} v_0  \\ 0 \end{array}\right) \right|^2 
\ee
where the $1/(N+1)$ comes from the zero-mode
normalization.  We can absorb it into renormalized
physical couplings, $g_1$ and $g_2$:
\be
{\cal{L}}_{Higgs} 
 \rightarrow 
\sum_{j=0}^{N} \left|(i{g}_2 A_{j, \mu} ^a\frac{\tau^a}{2}+i{g}_1 A_{j,\mu} \frac{Y}{2})
 \left( \begin{array}{c} v_0  \\ 0 \end{array}\right) \right|^2 
\ee
These terms may be rewritten in term of $W$, $Z$ and $\gamma$
fields on each brane:
\bea
{\cal{L}}_{Higgs} 
&= & \sum_{j=0}^{N} 
 M_{W}^2 W_{j\mu}^+W^{j\mu}{}^- + \half M_{Z}^2 Z_{j\mu} Z^{j\mu} 
\eea
The $W_i$ and $Z_i$ fields are combined with the Nambu-Goldstone
bosons $\pi^a$. The combined fields are defined as:
\begin{eqnarray}
W^\pm_{j\mu} & = & (A_{j,\mu}^1 \pm iA_{j,\mu}^2)/\sqrt{2}
\nonumber \\
{\gamma}_{j,\mu} & = & \sin\theta\; {A}^3_{j,\mu} + \cos\theta\; {A}_{j,\mu}
\nonumber \\
{Z}_{j,\mu} & = & \cos\theta\;  {A}_{j,\mu}^3 - \sin\theta\;  {A}_{j,\mu}
= \frac{(\tilde{g}_2  {A}_{j,\mu}^3 
- \tilde{g}_1  {A}_{j,\mu})}{\sqrt{\tilde{g}_1^2 + \tilde{g}_2^2}}
\end{eqnarray}
where $\gamma_{j,\mu}$ is a photon field, while $Z_j$ $(W_{j,\mu})$ is
a $Z$-boson ($W$-boson) mode. 

The masses $M_W$ and $M_Z$ are universal to all
the $SU(2)\times U(1)$'s, i.e., to all
branes,  and they are just the masses of the $W$ and $Z$ measured
in the low energy theory: 
\begin{eqnarray}
M_W &=& \frac{\tilde{g}_2^2 v_0^2}{2} = \frac{g_2^2 \tilde{v}_0^2}{2} \\
M_Z &=& \frac{(\tilde{g}_2^2 + \tilde{g}_1^2)v_0^2}{2} 
= \frac{(g_2^2 + g_1^2) \tilde{v}_0^2}{2},  
\end{eqnarray}
where $g_1$, $g_2$ and $\tilde{v}_0$ are measured at low energies. 

Combining these expressions with the full KK mass
formula, we find that the $W$, $Z$ and $\gamma$ KK
towers are given by:
\be
M^n_\gamma{}^2 =  4 M_0^2\sin^2\frac{\gamma_n}{2}
\ee
\be
M^n_W{}^2 =  M_W^2 + 4 M_0^2\sin^2\frac{\gamma_n}{2}
\ee
\be
M^n_Z{}^2 =  M_Z^2 + 4 M_0^2\sin^2\frac{\gamma_n}{2}
\ee
Each of the KK mode levels thus has a fine-structure determined
by the electroweak symmetry breaking.

\section{Incorporating Fermions}

\subsection{Chiral Fermions}

In $4+1$ dimensions free
fermions are vectorlike.
Chiral fermion zero modes can be
readily engineered.  For example, 
one can use domain
wall kinks in a background
field  which couples to  the fermion 
like a mass term. This can trap a chiral zero-mode
on the kink \cite{kaplan}. 
The magnitude of the kink field
away from the domain wall can be arbitrarily
large, so the vectorlike fermion masses
can be made arbitrarily large, and
are not directly related to the compactification scale.
This means that we need be concerned at present {\em only}
with the chiral zero-modes.
That is, from the point of view of our $3+1$
effective Lagrangian approach,
if we are only interested in the 
fermionic
zero modes then we can simply incorporate
the chiral fermions by hand.  

Consider one complete generation of left-handed
quarks and leptons, $\ell_L$, $q_L$ which are doublets
under the specific $SU(2)_{jL}$ and carrying weak
hypercharges $Y_\ell = -1$, $Y_{q}= 2/3$
under the $U(1)_{jY}$; the quarks carry
color under $SU(3)_j$; the fermions are sterile under
all other gauge groups $i\neq j$.  Likewise, we have
right-handed 
$SU(2)$ singlets, $\ell_{R}$, $q_{uR}$, and
$q_{dR}$ carrying 
weak hypercharges under the $U(1)_{jY}$.  
Additional generations can be incorporated with additional
fields. 

The chiral fermions of a given
generation can be placed
at a unique brane, distinct from the others.  One could
go further and split members within a single
generation.
In a sense this latter approach would emulate
the split-fermion construction
of Arkani-Hamed and Schmaltz, \cite{schmaltz}.
It leads us into interesting issues involving anomalies,
and Wess-Zumino terms in the present formulation
which we prefer to address elsewhere.
We will emulate more closely the
split family model \cite{dvali},
as we will presently consider a complete anomaly free
generation on any given brane.

Let us designate
the branes which receive the generations by $j=(j_1, j_2, j_3)$,
thus the full fermionic Lagrangian becomes:
\be
\label{fermion}
{\cal{L}}_{fermion} =  
\sum_{j}\left( \bar{\ell}_{j,L}\slash{D}_{j}\ell_{j,L}
+ \bar{q}_{j,L}\slash{D}_{j}q_{j,L} +
\bar{\ell}_{j,R}\slash{D}_{j}\ell_{j,R}
+ \bar{q}_{j,uR}\slash{D}_{j}q_{j,uR} + \bar{q}_{j,dR}\slash{D}_{j}q_{j,dR}
\right)
\ee
where $\slash{D}_{j} = \gamma^{\mu}(\partial_{\mu}-i\tilde{g}_2 A_{j, \mu}
^a\frac{\tau^a}{2} - i \tilde{g}_1 A_{j,\mu} \frac{Y}{2})$,
and the sum extends over  $j=(j_1, j_2, j_3)$.
The
couplings to the zero--mode gauge boson of
e.g., the quarks, are therefore 
\be
\label{fermion2}
{\cal{L}}_{0} =  \sum_{j}\left(\bar{q}_{j,L}\slash{\tilde{D}}q_{j,L}
+ \bar{q}_{j,uR}\slash{\tilde{D}}q_{j,uR} +
\bar{q}_{j,dR}\slash{\tilde{D}}q_{j,dR} \right)
\ee
where $\slash{\tilde{D}} = \gamma^{\mu}
(\partial_{\mu}-ig_2 \tilde{A}_{0, \mu}^a\frac{\tau^a}{2} 
- i g_1 \tilde{A}_{0,\mu} \frac{Y_{\psi}}{2})$, in
which $g_1$ and $g_2$ are the 
physical gauge coupling constants.

In the preceding discussion we considered a universal
Higgs field in the bulk.  This translated
into $N+1$ Higgs fields, $H_i$
each transforming as $\left(\half_i\right)$ (and singlet under
$j\neq i$, and with weak hypercharges $Y_i=1$ 
and $Y_j=0$ $j\neq i$.  This led to 
the zero mode gauge fields feeling a Higgs VEV
of order $m_H^2/\lambda \sim (N+1)m_H^2/\tilde{\lambda}$,
which is the conventional Standard Model result where
$\lambda$ is the physical (renormalized) low energy
quartic coupling.
Hence, one requires 
a tiny and unnaturally small Higgs boson
mass, $ m_H$ to generate the electroweak
symmetry breaking scale. 
The power law running of the coupling $\tilde{\lambda}$ brings
$\tilde{\lambda}$ at the
fundamental high evergy scale ($M_s$) down
to a low scale
$\lambda = \tilde{\lambda}/(N+1)$. To match on to the measured EW theory, 
one requires the mass-squared in the Higgs potential $m_H^2 \lta v_0^2$
which may be viewed as the present electroweak radiative bound,
whence $\lambda\lta 1$. If one saturates perturbative
unitarity and assumes
$\tilde{\lambda} \sim 16\pi^2$ at $M_s$, then the
KK tower is bounded by $N\lta 16\pi^2$.  

We would have expected that
the natural scale
for the Higgs mass is of order
the fundamental scale of the theory, $M_s$.
Can we modify the approach to introducing the Higgs
in such a way that the light Higgs boson becomes natural?
For example, can we engineer a Higgs mass of order
$M^2_s/N$ by judicious choice of the structure of the
model? 

One possibility is to assume that the Higgs potential
is non-universal, i.e., takes different values of
it's parameters for different values of $j$. The simplest
idea is to assume that a single Higgs on the $k$th brane
has a large negative mass-squared $\sim - m_H^2$ and the Higgs
gets a VEV on that brane only. 
This helps considerably, but does not alleviate the
naturalness problem. If $\VEV{H_k}\sim v$ then we get
a gauge mass term  $\tilde{g}^2(A_k)^2 v^2$ where $k$ is
unsummed.  However $A_k = A_0/\sqrt{N} + ...$ so again
the zero-mode mass term becomes   $\tilde{g}^2(A_0)^2 v^2/N$
 $\sim$ ${g}^2A_0^2v^2$. This requires
 that $v=v_0$, which implies that
 on the $k$th brane the Higgs mass
 is given by $v_0^2 = m_H^2/\tilde{\lambda}$.
 Note that now there is no large $(N+1)$ prefactor.
Using perturbative unitarity for $\tilde{\lambda}\lta 16\pi^2$,
we have an upper limit on $m_H\sim 1$ TeV 
(the Lee-Quigg Thacker bound \cite{Lee}).
Thus, this localization of the Higgs allows us
to raise the scale of the Higgs boson somewhat.
However, given that we typically want $N>>1$ we require $m_H<< M_s$,
so again we have an unnatural situation.
These are the two extreme limits of a zero-momentum VEV
and a localized (all momentum) VEV.

Despite the fact that the fundamental Higgs field
is unnatural in these schemes, it
is interesting to examine a latticized 
version of the split-generation model. Thus
we consider a model in which there is
a strongly localized Higgs VEV  \cite{dvali}.
We assign the Higgs VEV $v_0$
only to the $0$th brane, then the zero mode
gauge fields acquire masses of order $\tilde{g}^2v_0^2/N
\sim g_2^2v_0^2$.

The Higgs VEV exponentially attenuates
away from the localization point
and fermions that are at various distances from
the localized VEV will receive different
values. 
We assume the same structure as in eq.(\ref{higgs1}
where now the Higgs potentials have an $i$-dependent
mass term: 
\be
V(H_i) = M_i^2 H_i^\dagger H_i + \frac{\lambda_i}{2} (H_i^\dagger H_i)^2
\ee
For concreteness as an explicit example we choose:
\be
M^2_{i=0} = -\kappa M^2\qquad
M^2_{i\neq 0} = +M^2 \qquad \lambda_{i\neq 0} = 0
\ee
$\kappa$ is a phenomenological parameter.
The
full Higgs-only potential can be written:
\be
\label{temp}
V_{Higgs} = -\tilde{M}^2 H_0^\dagger H_0 
+ \frac{\lambda}{2} (H_0^\dagger H_0)^2
+ \sum_{i=1}^N \Lambda^2 H_i^\dagger H_i 
- \sum_{i=0}^N (M_0^2H_{i+1}^\dagger H_i + h.c.)
\ee
where we identify $H_{N+1} = 0$
($H_{N+1} = H_0)$ in the aliphatic (periodic) case
and thus
\be
\tilde{M}^2 = \kappa M^2 - M_0^2
\qquad \Lambda^2 = M^2 +2M_0^2
\ee
The equation of motion of the $H_i$ is thus:
\be
\Lambda^2 H_i = M_0^2H_{i+1} + M_0^2 H_{i-1}  \qquad (i \geq 1)
\ee
which has the solution $H_{i+1} = \epsilon H_i$ where:
\be
\epsilon = \frac{\Lambda^2 - \sqrt{\Lambda^4 - 4M_0^4 }}{2M_0^2}
\ee
If we substitute the solution back into the action of
eq.(\ref{temp}) we see that we obtain:
\be
\label{temp}
V_{Higgs} = -\tilde{M}^2 H_0^\dagger H_0 
+ \frac{\lambda}{2} (H_0^\dagger H_0)^2  -M_0^2 H_0^{\dagger}H_1,
\ee
and we can thus minimize the potential on
the zeroth brane as:
\begin{eqnarray}
\VEV{\tilde{H}_0} = \left( \begin{array}{c} v_0  \\ 0
\end{array}\right) 
\end{eqnarray} 
where $v_0^2 = M_H^2/\lambda$, $H_1=\epsilon H_0$
and $M_H^2 = \kappa M^2
-M_0^2 +\epsilon M_0^2$.

We can substitute the full dynamical Higgs
field into this expression, 
\begin{eqnarray}
{\tilde{H}_n} = \left( \begin{array}{c} v_{0n}+h_n/\sqrt{2}  \\ 0
\end{array}\right) 
\end{eqnarray}
and we have:
\be
v_{0n} = \epsilon^nv_0 \qquad h_n = \epsilon^nh_0
\ee
Now, we substitute into the kinetic terms of
eq.(\ref{higgs2}) to obtain the dynamical Higgs
field kinetic term:
\be
\sum_{n=0}^{N} (D_\mu H_i)^\dagger (D^\mu H_i)
\rightarrow \half[1 + \sum_{n=1}^{N} \epsilon^{2n}](\partial h)^2
\ee
We see that the dynamical
Higgs field has a wave-function renormalization constant:
\be
Z = [1 + \sum_{n=1}^{N} \epsilon^{2n}] = \frac{1-\epsilon^{2N+2}}{1-\epsilon^{2}}
\ee
Thus, the physical mass of the Higgs field becomes:
\be
m_H^2 = 2 M_H^2/Z
\ee
The Higgs is strongly localized in the limit
$M_0^2/\Lambda^2 \rightarrow 0$. In this limit $\epsilon\rightarrow 0$
and the only Higgs field receiving the VEV is effectively
$H_0$.  Then the zero-mode gauge masses are given by 
$\propto \tilde{g}^2v_0^2/{(N+1)} \sim {g}^2v_0^2$ and we
see that $v_0$ is indeed the electroweak VEV. Since 
$v_0^2 \sim  M_H^2/\lambda$ 
we see that $M_H\lta 1$ TeV,
by perturbative unitarity,  $\lambda \lta 16\pi^2 $.
We furthermore see that the physical Higgs is heavy, as 
$ m_H^2 \sim 2 M_H^2/Z \sim $ TeV. In this case, $\epsilon \sim
M_0^2/\Lambda^2 \ll 1$ implies that $M^2 \gg M_0^2$. The most natural way to
generate the EW scale $M_H^2$ is thus to 
tune a cancellation between $\kappa M^2$ and $-M_0^2$ and
use small $\epsilon$ to account for the hierachy between $M_0^2$ and the EW
scale. 

On the other hand, we can delocalize the Higgs with
$\epsilon \rightarrow 1 - \eta$ and $\eta << 1$. Then
we see that $Z\rightarrow (N+1)$. Now 
the zero-mode gauge masses are given by 
$\propto \tilde{g}^2v_0^2 \sim (N+1){g}^2v_0^2$ and we
see that $\sqrt{N+1}v_0$ is the electroweak VEV. 
This recovers the universal Higgs configuration described
in Section 3.

\subsection{Localization and the Split-Generation Model}

Restoring the link-Higgs
fields for gauge covariance, 
the nearest neighbor interactions generates a profile
for the Higgs field
of the form 
$H_j = \Pi_{i=0}^{j} (\epsilon \Phi_i{}' \phi_i{}'^3/v_2 v^3_1 ) H_0$, 
which is the
discretized version of the exponential attenuation in $x^5$ away
from the source $H(x^5)\sim \exp(-M |x^5|) H(0)$.

For diagonal masses we consider only the
fermions placed on a given brane.
If there is a complete family of fermions 
on the $j$th brane, it is charged under $SU(3)_j \times
SU(2)_j \times U(1)_j$ only.
We postulate a coupling to the Higgs field $H_j$ as: 
\be
{\cal{L}}_{Yukawa} =  y_{\ell j}\bar{\ell}_{j,L}H^c_j\ell_{j,uR} +
y_{uj}\bar{q}_{j,L}H_jq_{j,uR} + y_{dj}\bar{q}_{j,L}H^c_jq_{j,dR}  + h.c.
\ee
($H^c$ is the charge-conjugated Higgs field).
These fermions thus acquire masses as $\VEV{H_j}$ becomes non-zero, 
\be
\rightarrow {\cal{L}}_{mass} =
y_{\ell j}v_0\epsilon^{j}\bar{\ell}_{j}\ell_{j} +
y_{uj}v_0\epsilon^{j}\bar{u}_{j}u_{j} +
y_{dj}v_0\epsilon^{j}\bar{d}_{j}d_{j} 
 \ee
If we place the three fermion generations on different branes
$j_1\neq j_2\neq j_3$,  
the 
diagonal hierachy between the families is generated through the 
suppression factors $\epsilon^{j_i}$ \cite{dvali}. 

The off-diagonal terms in the mass matrix must
be generated to give a nontrivial CKM matrix.
We specialize to quarks. This mixing
now arises through higher dimensional
operators corresponding to 
the overlap of the wave-functions
of the chiral zero-mode fermions localized on different branes: 
\be
{\cal{L}}_{mixed} = y_{u,il}\bar{q}_{j_i,L} H_{j_i} \left( \Pi_{l=j_i+1}^{j_l}
\frac{\bar{\phi}_l^4}{M_f^4} \right) q_{j_l,uR} +  y_{d,il}
\bar{q}_{j_i,L} H^c_{j_i} \left( \Pi_{l=j_i+1}^{j_l}
\frac{\phi_l^2}{M_f^2} \right) q_{j_l,dR} 
\ee
where the fields $\tilde{\Phi}_l$ are composites
defined as $\Phi^{'}_l \phi^{'}_l$.
We emphasize that the
mass scale $M_f$ is new, and
is related to the masses of the decoupled 
vectorlike fermions. The above expression effectly
mimics the overlapping of fermion wave functions in the set-up of split
fermions \cite{schmaltz}, \cite{dvali}. The suppressed off-diagonal mass terms are
therefore: 
\be
{\cal{L}}_{mixed} =
y_{u,il}v_0(\epsilon')^{4|j_n-j_i|}\epsilon^{j_i}\bar{u}_{j_i,L}u_{j_l,R} 
+  y_{d,il}v_0(\epsilon')^{2|j_n-j_i|}\epsilon^{j_i}
\bar{d}_{j_i,L} d_{j_l,R} + h.c.,  
\ee
where $\epsilon{'} = v/M_f$. In this manner a model
of the CKM matrix can be generated.

We will not presently address the effective Lagrangian
and the phenomenology of
the split generations in detail
at present, in particular
the problematic coupling to the KK modes.  As a consequence
of splitting, this is non-universal and flavor-changing
neutral current effects occur \cite{poma}.  One can
live with these by raising the compactification
mass scale. 
Of course, at the end of the day we may view this as
a $3+1$ dimensional model in which there are many
mixing interactions and higher dimension
operators giving the hierarchy.  Perhaps we
can discover new GIM symmetries to suppress
such effects.

\section{Discussion and Conclusion}

In conclusion, we have given a description
of the Standard Model in the bulk as a pure $3+1$ dimensional
effective theory.  One can in principle discard
the notion of an extra-dimension and view this
as an extension of the Standard Model
within $3+1$ dimensions with extra discrete symmetries.
The connection to extra dimensions is made through
the transverse lattice, and this may be viewed as a 
manifestly gauge invariant
low energy effective theory for an extra-dimensional
Standard Model.  Softening the link-Higgs fields to
dynamical Higgs fields leaves a renormalizable 
effective Lagrangian (modulo certain higher
dimension operators that are involved in fermion
mass and mixing angle physics).

The larger gauge invariance needed to describe
KK modes in $3+1$ may be viewed as a consequence of
hidden local symmetries required to make renormalizable
theories of spin-1 objects \cite{hidden}.  Alternatively,
this is the expanding local gauge invariance 
in the bulk that appears as an
extra dimension opens up.

In treating the $\phi$ weak hypercharge link Higgs
fields we have, strictly speaking, departed somewhat
from the pure transverse lattice.  In the chiral phase
we could have used fractional powers of a $\phi$
link with $Y=1$ to propagate quarks, but we chose the
present decomposition to maintain a polynomial
effective Lagrangian.

We do not, alas, gain insights into the
problem of naturalness of the Higgs mass and
electroweak hierarchy. 
Many issues remain, however, to be addressed in the context of the
general
transverse lattice approach to describing extra dimensions
\cite{hill}.  For example, how does a dynamical
electroweak symmetry breaking scheme emerge in this
description \cite{dobrescu}?  One thing we see immediately
in this approach is the emergence of an imbedding
of QCD as in $SU(3)\rightarrow SU(3)\times SU(3)$, etc.
This is remniscent of the structure
of Topcolor, \cite{topc}, and suggests
that class of extra-dimensional 
models in which the electroweak symmetry
is broken dynamically \cite{dobrescu}.

We view the transverse lattice
approach as providing powerful new insights
into the construction of new extensions beyond the 
Standard Model within $3+1$ model building. Many
future applications to SUSY, gravity, topology,
strong dynamics, and grand
unification are foreseeable.

\vspace*{1.0cm}
\noindent
{\bf \Large \bf Acknowledgements}

We wish to thank W. Bardeen, E. Eichten and
M. Schmaltz for useful discussions.
One of us (SP) wishes acknowledge the kind hospitality of the
Fermilab theory group where this work originated.
H.-C. Cheng is supported by the Robert R. McCormick Fellowship and by
DOE grant DE-FG02-90ER-40560. 
Research by CTH and JW was supported by the U.S.~Department of Energy
Grant DE-AC02-76CHO3000.
Research by SP was supported by the Polish State Committee
for Scientific Research, grant KBN 2 P03B 060 18
(2000-01).
\frenchspacing

\end{document}